\documentclass[prl,amssymb,twocolumn,superscriptaddress,showpacs]{revtex4}

\usepackage{graphicx}

\begin{document}

\title{Non-linear photonic crystals as a source of entangled photons}

\author{Michiel J.A. de Dood} \email[corresponding author: ]{mdedood@physics.ucsb.edu}
\affiliation{Department of Physics, University of California Santa Barbara, CA 93106, Santa Barbara, USA}

\author{William T.M. Irvine}
\affiliation{Department of Physics, University of California Santa Barbara, CA 93106, Santa
Barbara, USA}
\affiliation{ Department of Physics, University of Oxford, Parks Road, Oxford
OX1 3PU, United Kingdom}

\author{Dirk Bouwmeester}
\affiliation{Department of Physics, University of California Santa Barbara, CA 93106, Santa Barbara, USA}


\begin{abstract}
Non-linear photonic crystals can be used to provide phase-matching for frequency conversion in optically
isotropic materials. The phase-matching mechanism proposed here is a combination of form birefringence and
phase velocity dispersion in a periodic structure. Since the phase-matching relies on the geometry of the
photonic crystal, it becomes possible to use highly non-linear materials. This is illustrated considering a
one-dimensional periodic Al$_{0.4}$Ga$_{0.6}$As / air structure for the generation of 1.5 $\mu$m light. We
show that phase-matching conditions used in schemes to create entangled photon pairs can be achieved in
photonic crystals.
\end{abstract}

\maketitle

Polarization entangled photon pairs play a central role in testing the foundations of quantum mechanics and
in implementing quantum information protocols~\cite{Bouwmeester:Book}. They are likely to remain an appealing
resource for practical quantum information science since they interact very little with the environment,
propagate easily over long distances and since photon polarization is easily manipulated in experiments. A
popular method to create such pairs is by down-conversion in non-linear crystals. For this process to be
efficient the down-converted photons generated by pump photons at different points in the crystal have to be
in phase with each other. This leads to the ``phase-matching'' condition $\vec{k}_p = \vec{k}_1 + \vec{k}_2$,
for the wavevectors of the pump ($\vec{k}_p$) and downconverted photons ($\vec{k}_1, \vec{k}_2$). In general
this condition will not hold owing to normal index dispersion which makes $|\vec{k}_p| > |\vec{k}_1| +
|\vec{k}_2|$. The current schemes for down-conversion employ the natural birefringence of specific non-linear
crystals, like $\beta$-Barium-Borate (BBO), to compensate for this effect.

In one of these schemes~\cite{Kwiat:PRL1995,Kurtsiefer:JMO2001}, down-converted photons emerging in a
particular pair of directions are entangled in polarization:
\begin{eqnarray}
|\Psi\rangle = \frac{1}{\sqrt{2}} (|V\rangle_1 |H\rangle_2 + \mathrm{e}^{i \theta} |H\rangle_1 |V\rangle_2 ),
\label{Eq:Entangled}
\end{eqnarray}
where $V$ and $H$ denote the polarization state of particles 1 and 2 and $\theta$ is a phase factor that can
be set by passing one of the photons through a phase-plate.

Although birefringent non-linear crystals have proven to be a successful source in many proof of principle
experiments the demands of both practical quantum information and schemes for the implementation of all
optical quantum computing~\cite{Knill:NAT2001} make it desirable to produce sources that are both more
efficient and more easily integrable on, for example, an optical chip. The requirement that both the
birefringence and the $\chi^{(2)}$ non-linearity be naturally present severely limits the possibilities for
improvement of sources based on existing crystals. Semiconductors such as GaAs or GaP have a $\chi^{(2)}$,
typically 200 pm/V~\cite{Bergfeld:PRL2003}, about two orders of magnitude larger than that of commonly used
crystals such as BBO, 2.2 pm/V~\cite{Yariv:Waves}. This, together with the existence of well developed
micro-fabrication techniques for these materials, makes it attractive to explore ways of creating
semiconductor based entangled photon sources. In these materials, which have no natural birefringence, the
conditions for phase-matching must be created artificially. In this letter we show how photonic crystals can
be used to achieve this.

Photonic crystals are materials with a periodic variation in refractive index on the scale of the optical
wavelength. In layered structures the different boundary conditions at the interfaces for the two
polarizations leads to an effective or ``form'' birefringence. The application of this effect to
phase-matching has been explored in the limit that the optical wavelength is much larger than the
periodicity~\cite{Ziel:APL1976,Fiore:NAT1998,deRossi:APL2001}. When the wavelength is comparable to the
periodicity it was suggested that a reciprocal lattice vector can be added to the phase-matching
conditions~\cite{Sakoda:PRB1996}. Alternatively a similar relation was found for periodically poled
structures that have a periodicity in $\chi^{(2)}$, leading to quasi-phase-matching
conditions~\cite{Berger:PRL1998}. Our work proposes how a combination of form birefringence and the strongly
altered phase velocity in a photonic crystal can be used to phase-match down-converted light and generate
entangled photons.

We focus on one-dimensional structures for sake of simplicity, although the concepts can be extended to two-
and three-dimensional photonic crystals. The periodic variation in refractive index leads to Bragg scattering
of the light and wave propagation becomes best described in terms of a photonic bandstructure. If the
wavelength is comparable to the periodicity of the structure, the propagation of light is strongly affected,
leading to the existence of a range of frequencies, known as a stopband, for which light does not propagate.

The propagation of electromagnetic waves in a photonic crystal is described by Maxwell's equations with a
periodic dielectric function $\epsilon(\vec{r})$. The general solution is given in terms of Bloch waves
labeled by a frequency $\omega$ and a Bloch-wave vector $\vec{K}$ and consists of a plane wave multiplied by
a function that has the periodicity of the photonic lattice. The Bloch wave can be written as a Fourier sum
over the reciprocal lattice vectors $\vec{G}$:
\begin{equation}
\vec{E}_{\vec{K}}(\vec{r},t)=\mathrm{e}^{-i(\vec{K} \cdot \vec{r}-\omega t)} \sum_{\vec{G}}
\vec{e}_{\vec{K},\vec{G}} \mathrm{e}^{-i\vec{G} \cdot \vec{r}},\label{Eq:Bloch}
\end{equation}
where $\vec{e}_{\vec{K},\vec{G}}$ are the Fourier coefficients for each of the space harmonics. For a
one-dimensional structure $\vec{K}$ and the coefficients $\vec{e}_{\vec{K},\vec{G}}$ can be obtained by
finding the eigenvalues of a $2\times2$ transfer matrix for each polarization separately~\cite{Yariv:Waves}.
The effect of index dispersion, essential in any discussion of phase-matching, can be easily incorporated as
the transfer matrix is defined at each frequency separately. The obtained wavevector can be decomposed in
components $K_z$ perpendicular and $k_{\parallel}$ parallel to the layers of the photonic crystal. The
dispersion relation for transverse electric (TE) or ordinary ($o$) polarized waves, that have the electrical
field in the plane of the layers, is given by~\cite{Yariv:Waves}:
\begin{eqnarray}
K_{z}(\omega, k_{\parallel}) = \frac{1}{\Lambda} \arccos[\cos(k_{1z} a + k_{2z} b) - \nonumber
\\ \frac{1}{2} \frac{(k_{1z}+k_{2z})^2}{k_{1z} k_{2z}} \sin(k_{1z} a)\sin(k_{2z} b)],
\label{Eq:Dispersion}
\end{eqnarray}
where $k_{1z}$ and $k_{2z}$ are the plane wavevector components perpendicular to the interfaces in medium one
(of refractive index $n_1$) and medium two (of refractive index $n_2$):
\begin{eqnarray}
k_{1,2z}=\sqrt{\bigg(\frac{n_{1,2}\ \omega}{c}\bigg)^2 - k_{\parallel}^2}. \nonumber
\end{eqnarray}
An  analogous expression can be derived for the orthogonal transverse magnetic (TM) or extraordinary ($e$)
polarization, that results in different propagation constants. This important polarization dependence arises
from the different boundary conditions at the interfaces for the two polarizations and will play a key role
in the generation of polarization entangled photon pairs.

\begin{figure}[tb!]
\includegraphics[width=86mm]{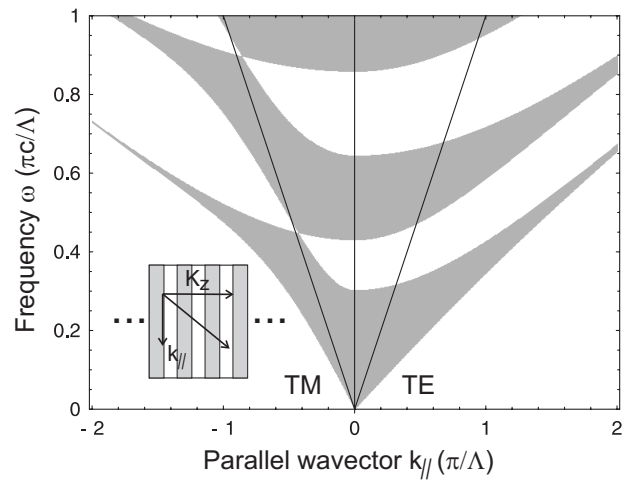}
\caption{Bandstructure of a periodic structure with Al$_{0.4}$Ga$_{0.6}$As and air layers for TM (left) and
TE (right) polarized light. The grey area corresponds to propagating solutions at a given frequency $\omega$
and wavevector $k_{\parallel}$. The solid lines indicate the ``light line'' in vacuum: $\omega = c
|\vec{k}|$. \label{Fig:bandstructure} }
\end{figure}

Figure~\ref{Fig:bandstructure} shows a bandstructure derived from Eq.~\ref{Eq:Dispersion} for a periodic
structure with alternating layers of Al$_{0.4}$Ga$_{0.6}$As and air. The fill fraction of
Al$_{0.4}$Ga$_{0.6}$As is 0.656. The gray area corresponds to propagating solutions for TM waves (left panel,
negative $k_{\parallel}$) and TE waves (right panel, positive $k_{\parallel}$). They coincide for normal
incidence ($k_{\parallel}$~=~0). The solid lines in the figure correspond light in vacuum ($\omega = c
|\vec{k}|$) and divide the modes in those accessible to waves from outside the crystal, and those that are
confined by total internal reflection. The latter modes can be accessed from the side of a sample or by using
a set of prisms~\cite{Tien:APL1969}.

The frequency is specified in multiples of $\pi c / \Lambda$, where $\Lambda$ is the periodicity of the
structure. The structures were designed by first considering the refractive index to be equal to that at the
pump frequency $\omega_p$ ($n$~=~3.4 at $\lambda$~=~750 nm). This allows the selection of a periodicity
$\Lambda$ such that the pump photons propagate in the structure. Once the periodicity is fixed, the frequency
dependent refractive index is used to calculate dispersion surfaces to search for phase-matching conditions
at relevant frequencies. The choice of material is motivated by the large second order non-linearity together
with the fact that it is transparent at a pump wavelength of 750 nm, to allow degenerate down-conversion in
the important telecommunication window around 1500 nm. Moreover, the fabrication of such crystals seems to be
feasible by starting with a periodic structure that has layers of AlAs and Al$_{0.4}$Ga$_{0.6}$As followed by
selective wet etching of the AlAs~\cite{Yablonovitch:APL1987}. The fill fraction was chosen in order to
maximize birefringence in the long wavelength limit ($\Lambda \ll \lambda$), where the Bloch waves are plane
waves and experience an effective medium that behaves as a uniaxial birefringent material.

The leading terms in the Bloch wave expansion in Eq.~\ref{Eq:Bloch} will phase-match when
$\vec{K_p}+\vec{G_p} = \vec{K_1} + \vec{G_1} +  \vec{K_2} + \vec{G_2}$ where the $\vec{G}$s correspond to the
leading $\vec{e}_{\vec{K},\vec{G}}$s, ensuring efficient down-conversion into these modes. Two regimes are
explored that lead to the generation of entangled photon pairs. We first discuss the long wavelength limit
($\Lambda \ll \lambda$) followed by the case where $\Lambda \sim \lambda$.

\begin{figure}[tb!]
\includegraphics[width=86mm]{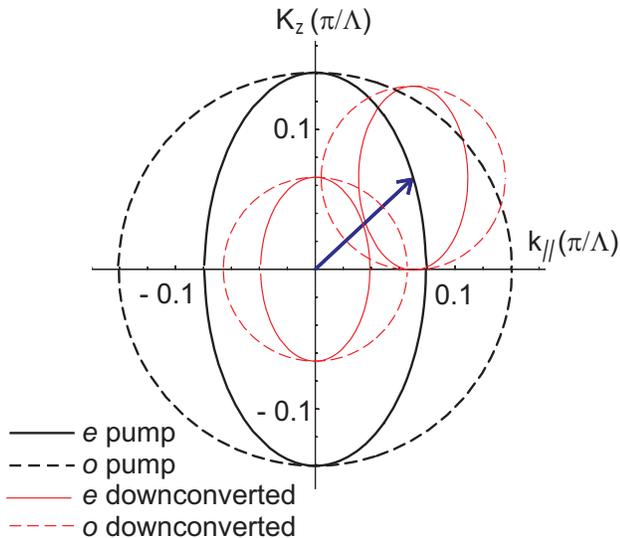}
\caption{Dispersion surfaces for degenerate down-conversion at 1500 nm in a one-dimensional
Al$_{0.4}$Ga$_{0.6}$As / air photonic crystal with a periodicity $\Lambda$~=~18.75~nm and a fill fraction of
Al$_{0.4}$Ga$_{0.6}$As of 65.6\%. Dispersion surfaces are shown for the pump at a wavelength of 750~nm (thick
lines) and for degenerate down-conversion at 1500 nm (thin lines). The dashed lines correspond to TE
(ordinary) polarization, while the solid lines correspond to TM (extraordinary) polarization. The pump
wavevector indicated by the arrow is extraordinary. \label{Fig:PhasematchLongWavelength} }
\end{figure}

Figure~\ref{Fig:PhasematchLongWavelength} shows dispersion surfaces in the long wavelength limit for a
structure with periodicity $\Lambda$~=~18.75~nm and a pump wavelength of 750~nm (0.05 $\pi c/\Lambda$). One
set of dispersion surfaces for the down-converted light, at 1500~nm wavelength, is drawn at the origin,
another is drawn at the end of the pump wavevector in order to obtain a simple geometric construction of the
phase-matching condition. Wavevectors satisfying the phase-matching condition correspond, in such a diagram,
to the intersections between the dispersion surfaces of the down-converted photons.

The intersections of the thin dotted lines represent the wavevectors of two ordinary ($o$) polarized photons,
created by the extraordinary ($e$) polarized pump field. This situation is usually referred to as
non-colinear type-I down-conversion. Since the structure is symmetric under rotations about the optical axis
parallel to $K_z$, the circles and ellipses in Fig.~\ref{Fig:PhasematchLongWavelength} should be seen as
spheres and ellipsoids. The intersections of the two down-conversion spheres for $o$ polarization will give a
circle, representing a cone of down-converted light centered on the pump wavevector. Every pair of photons is
emitted in such a way that they obey the phase-matching condition and so are always diametrically opposite
each other about the pump.

For noncolinear type-II down-conversion, the pump field is $e$ polarized and the down-converted photon pairs
each have one $o$ polarized and one $e$ polarized photon. This case corresponds to the intersections of the
dashed thin lines with the solid thin lines in Fig.~\ref{Fig:PhasematchLongWavelength}. There are two pairs
of such intersections, each representing an intersection between a sphere and an ellipsoid. They define two
slightly distorted cones of down-converted light whose central axes are not aligned. For the direct
generation of polarization entangled photons as in Eq.~\ref{Eq:Entangled}, the $e$ and $o$ photons in a pair
must be emitted in directions in which $e$ photons and $o$ photons are indistinguishable except by their
polarization. This happens at the intersections of the two distorted cones discussed above.

\begin{figure}[tb!]
\includegraphics[width=86mm]{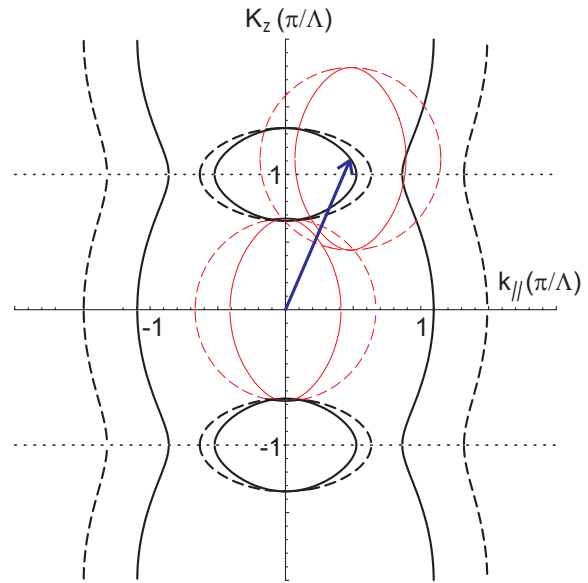}
\caption{Dispersion diagram in the repeated zone scheme for a pump wavelength of 750 nm. The periodicity of
the crystal is $\Lambda$~=~187.5~nm and the fill fraction of Al$_{0.4}$Ga$_{0.6}$As is $\alpha$~=~0.656. The
pump wavevector is indicated by the arrow. \label{Fig:TypeII}}
\end{figure}

Figure~\ref{Fig:TypeII} shows dispersion surfaces for a photonic crystal with a periodicity
$\Lambda$~=~187.5~nm, a factor of 10 greater than that of Fig.~\ref{Fig:PhasematchLongWavelength} and
comparable to the wavelength of the pump light. The frequency of the pump $\omega$ = 0.5 $\pi c / \Lambda$
($\lambda$ = 750 nm) is now above the first stopband for normal incidence. The dispersion surfaces for the
pump are no longer continuous as some directions are excluded by Bragg reflection.

For the downconverted photons at a wavelength of 1500 nm, the Bloch waves are essentially plane waves with
wave vectors $\vec{K}_1$ and $\vec{K}_2$. The phase-matching condition in this case reduces to: $\vec{K}_{p}
+ \vec{G_p} = \vec{K}_{1} + \vec{K}_{2}$. The pump wavevector, indicated by the arrow is chosen to end in the
second Brillouin zone in order to select the reciprocal lattice vector that corresponds to the dominant space
harmonic of the Bloch wave in Eq.~\ref{Eq:Bloch}. In principle, phase-matching can also be achieved by
choosing another reciprocal lattice vector, but the amplitude of the down-converted light will be reduced.

The dispersion surfaces of down-converted photons in Fig.~\ref{Fig:TypeII} indicated by the thin lines
resemble those of the long wavelength limit. Again there are two different types of intersections,
corresponding to Type-I ($e \rightarrow o o$) and Type-II ($e \rightarrow e o$) down-conversion. As discussed
in the long wavelength limit, polarization entangled photons may be collected at the intersections of the
distorted cones.

In the long wavelength limit the phase-matching relies solely on form birefringence. However, when the
periodicity of the crystal is comparable to the wavelength of the pump, the phase-matching also relies on a
substantial change in phase velocity. For frequencies below the stopband the phase velocity is lower than
that in the long wavelength limit, which works against phase-matching since it adds to the effect of normal
index dispersion. However, if the frequency is above the frequency of the stopband the phase velocity is
increased and compensates the effect of index dispersion. For large enough index contrast this can be used to
achieve phase-matching where all photons have the same polarization, i.e. $e \rightarrow e e$ or $o
\rightarrow o o$.

We will now estimate the efficiency of our scheme compared to that of the scheme based on BBO. There are four
factors that contribute to the relative amplitude: (i) The ratio of the value of the $\chi^{(2)}$'s. (ii) The
efficiency per unit length should be multiplied by the fill fraction of $\chi^{(2)}$ material (0.65). (iii)
The scheme phase matches the leading term of the Bloch waves. The magnitude of this term is obtained by
Fourier analysis of the Bloch wave. This gives a factor of $\sim$1 and a factor $\sim$0.21 for the structures
in Fig.~\ref{Fig:PhasematchLongWavelength} and Fig.~\ref{Fig:TypeII} respectively. (iv) As for any non-linear
material we need to consider the tensor properties of the non-linearity. Al$_{0.4}$Ga$_{0.6}$As has the
$\bar{4}3m$ point group symmetry and for the conventional (100) surface orientation and our directions this
gives a factor of $\sim$0.83 and $\sim$0.53. Multiplying all factors and squaring to get the relative
efficiency gives $\sim$2500 and $\sim$50 times the efficiency for BBO.

The example structures we used to illustrate our scheme have by no means been tailored to maximize the
efficiency. The large value of $\chi^{(2)}$ in Al$_{0.4}$Ga$_{0.6}$As leaves ample room for tailoring the
geometry of the structure to suit a particular application, whilst keeping the process efficient. It should
be noted that there is a trade-off between increasing the strength of the interaction of light with the
photonic crystal and the loss in efficiency associated with the undesired terms in the Bloch wave expansion
(c.f. the efficiency for Fig.~\ref{Fig:PhasematchLongWavelength} and Fig.~\ref{Fig:TypeII}).  An interesting
extension of our work is to consider downconversion of pulsed gaussian beams, which is accompanied by
spatio-temporal walk-off effects that affect the quality of entanglement and the efficiency of the process.
It seems promising to explore the difference between group velocity and phase velocity in a photonic crystal
to reduce these effects.

The concepts introduced in this letter can be extended to two- and three-dimensional photonic crystals. The
change in phase velocity is a result of Bragg reflection common to all photonic crystals. Form birefringence
relies on polarization dependent boundary conditions and is observed in one and two-dimensional photonic
crystals. Form birefringence can be introduced in three-dimensional structures by making the individual
scatterers elongated which can be achieved in microfabricated structures~\cite{Lin:NAT1998} or by deforming a
colloidal crystals~\cite{Velikov:APL2002}.

We presented a phase-matching scheme that uses the combined effects of form birefringence and a change in
phase velocity in photonic crystals to enable the production of  entangled photon pairs. Both effects can be
tuned by designing an appropriate structure and are decoupled from the intrinsic non-linearity that is
determined by the choice of the constituent materials. We focused primarily on III-V semiconductors and on
AlGaAs in particular. These materials have a high non-linearity together with a large linear refractive index
which contribute to a strong interaction with light. Microfabrication techniques are well developed for these
materials and an experimental realization seems feasible. If successful, the structure could be implemented
on optical chips as an integrated source of  entangled photons.

We are most grateful to Mark Sherwin and Evelyn Hu for stimulating discussions.  W.I. acknowledges Elsag
s.p.a. for support under MIUR's grant no. 67679/L.488/92. M.d.D and D.B. acknowledge support under DARPA
grant no. MDA972-01-1-0027.

%% \bibliography{Entangled_Bragg}

\end{document}